\documentclass[aps,pre,onecolumn,groupedaddress,longbibliography,nofootinbib,showkeys,showpacs]{revtex4-1}

\usepackage{amsmath}
\usepackage{amsfonts}
\usepackage{graphicx}
\usepackage{mathptmx}      
\usepackage{bm}
\usepackage{enumerate}
\usepackage{subfigure}
\usepackage{color}
\usepackage{grffile}
\usepackage{grffile}

\newcommand\ba{\mathbf{a}}
\newcommand\bb{\mathbf{b}}
\newcommand\bc{\mathbf{c}}
\newcommand\bd{\mathbf{d}}
\newcommand\bu{\mathbf{u}}
\newcommand\bx{\mathbf{x}}

\newcommand\bp{\mathbf{p}}
\newcommand\bq{\mathbf{q}}

\newcommand*\widebar[1]{%
   \hbox{%
     \vbox{%
       \hrule height 0.5pt 
       \kern0.5ex
       \hbox{%
         \kern-0.1em
         \ensuremath{#1}%
         \kern-0.1em
       }%
     }%
   }%
}

\definecolor{DarkGreen}{RGB}{0,100,0}
\begin{document}


\title{Discrete-event simulation of an extended Einstein-Podolsky-Rosen-Bohm experiment}

\author{H. De Raedt}
\affiliation{Institute for Advanced Simulation, J\"ulich Supercomputing Centre,\\
Forschungszentrum J\"ulich, D-52425 J\"ulich, Germany}
\affiliation{Zernike Institute for Advanced Materials,\\
University of Groningen, Nijenborgh 4, NL-9747 AG Groningen, The Netherlands}
\author{M.S. Jattana}
\affiliation{Institute for Advanced Simulation, J\"ulich Supercomputing Centre,\\
Forschungszentrum J\"ulich, D-52425 J\"ulich, Germany}
\affiliation{RWTH Aachen University, D-52056 Aachen, Germany}
\author{D. Willsch}
\affiliation{Institute for Advanced Simulation, J\"ulich Supercomputing Centre,\\
Forschungszentrum J\"ulich, D-52425 J\"ulich, Germany}
\author{M. Willsch}
\affiliation{Institute for Advanced Simulation, J\"ulich Supercomputing Centre,\\
Forschungszentrum J\"ulich, D-52425 J\"ulich, Germany}
\author{F. Jin}
\affiliation{Institute for Advanced Simulation, J\"ulich Supercomputing Centre,\\
Forschungszentrum J\"ulich, D-52425 J\"ulich, Germany}
\author{K. Michielsen}
\affiliation{Institute\ for\ Advanced Simulation, J\"ulich Supercomputing Centre,\\
Forschungszentrum J\"ulich, D-52425 J\"ulich, Germany}
\affiliation{RWTH Aachen University, D-52056 Aachen, Germany}

\date{\today}

\begin{abstract}
We use discrete-event simulation to construct a subquantum model that can reproduce
the quantum-theoretical prediction for the statistics of data produced by the
Einstein-Podolsky-Rosen-Bohm experiment and an extension thereof.
This model satisfies Einstein's criterion of locality and generates data in an event-by-event
and cause-and-effect manner.
We show that quantum theory can describe the statistics of the simulation data
for a certain range of model parameters only.

\end{abstract}


\maketitle

\section{Introduction}\label{INT}

The Einstein-Podolsky-Rosen thought experiment was introduced to question the completeness
of quantum theory~\cite{EPR35}, ``completeness'' being defined in Ref.~\onlinecite{EPR35}.
Bohm proposed a modified version that employs the spins instead of coordinates and momenta of
a two-particle system~\cite{BOHM51}, and is experimentally realizable~\cite{KOCH67,CLAU78,ASPE82b,WEIH98,CHRI13,HENS15,GIUS15,SHAL15}.
A key issue in the foundations of physics is whether there exist ``local realist'' models
that yield the statistical results of the quantum-theoretical description
of the Einstein-Podolsky-Rosen-Bohm (EPRB) experiment.

In this paper, we take, as operational definition of a local realist model,
any model for which
\begin{enumerate}
\item
all variables, including those representing events which occur at
specific locations and specific times,
always have definite values,
\item
all variables change in time according to an Einstein-local, causal process.
\end{enumerate}

{\color{black}In the literature, one often finds the statement}
that Bell's theorem~\cite{BELL64,BELL93}
rules out {\bf any} local realist model for the EPRB experiments.
In Refs.~\onlinecite{BELL64,BELL93}, Bell gives a proof that a correlation ${C}(\ba,\bb)$ of the form
\begin{eqnarray}
{C}(\ba,\bb)&=&\int d\lambda \;\mu(\lambda) {A}(\ba,\lambda){B}(\bb,\lambda)\;,\quad
|{A}(\ba,\lambda)|\le1\;,\quad|{B}(\bb,\lambda)|\le1
\;,\quad 0\le\mu(\lambda)
\;,\quad\int d\lambda \;\mu(\lambda)=1
\;,
\label{IN0}
\end{eqnarray}
cannot arbitrarily closely approximate the correlation $-\ba\cdot\bb$
{\it for all} unit vectors $\ba$ and $\bb$.
According to Bell (see Ref.~\onlinecite{BELL93}), this is the theorem.
On the other hand, the quantum-theoretical description of the EPRB experiment
in terms of two spin-1/2 particles in the singlet state yields the correlation $-\ba\cdot\bb$.
Clearly, there is a conflict between the
quantum-theoretical model of the EPRB experiment and the model defined by Eq.~(\ref{IN0}).
While there can be no doubt about the mathematical correctness of Bell's theorem,
the physical relevance of the theorem and
its applicability to the data gathered in laboratory EPRB experiments
has been under scrutiny
since its conception~\cite{PENA72,FINE74,FINE82,FINE82a,FINE82b,MUYN86,KUPC86,BRAN87,JAYN89,BROD93,PITO94,FINE96,KHRE09z,
SICA99,BAER99,HESS01a,HESS01b,HESS05,ACCA05,KRAC05,SANT05,
KUPC05,MORG06,KHRE07,ADEN07,NIEU09,MATZ09,KARL09,KHRE09,GRAF09,KHRE11,NIEU11,HESS15,KUPC16z,KUPC17,HESS17a,NIEU17}.
A fundamental problem with the application of Bell's model Eq.~(\ref{IN0}) to this data  is the following.

Evidently, in a laboratory EPRB experiment, before one can even think about computing correlations of particle properties,
it is necessary to first classify a detection event as corresponding to the arrival of a particle or as something else.
Any laboratory EPRB experiment with photons employs a specific, well-defined procedure
to identify photons~\cite{KOCH67,CLAU78,ASPE82b,WEIH98,CHRI13,HENS15,GIUS15,SHAL15}.
Such a procedure is {\bf definitely missing} in the model Eq.~(\ref{IN0}) proposed and analyzed by Bell~\cite{BELL93}.
If the aim is to describe the outcome of a laboratory EPRB experiment,
then not incorporating such a procedure in the model is a fallacy which, logically speaking, is not much different
from trying to model electrodynamics in terms of electrical phenomena
without taking into account the magnetic phenomena.
Although it is good practice to analyze the most primitive model first,
the observation that it does not agree with experimental results only suggests that it is {\bf too primitive}.
The failure of the primitive model to account for the identification process is the fundamental  reason why Bell's theorem
cannot have the status of a ``no-go'' theorem for the existence of a local realist model for a laboratory EPRB experiment.

In this paper, we use the term ``subquantum model'' to refer to a local realist model of an experiment
which satisfies the requirements 1 and 2 and

\begin{enumerate}
\item[3.]
the model can reproduce the statistical results of the quantum-theoretical description
of the experiment in an event-by-event, cause-and-effect manner.
\end{enumerate}

The main aim of this paper is to present a subquantum model for the EPRB experiment and an extended version thereof.
{\color{black}The latter, which we abbreviate by EEPRB, differs from the standard EPRB experiment in that all the measurements for the four different pairs of settings, required to perform Bell-inequality tests, can be made in one single run instead of four runs of the experiment. As such, the EEPRB experiment is not vulnerable to the contextuality loophole~\cite{NIEU11}.
We adopt the discrete-event simulation (DES) approach, introduced in Ref.~\onlinecite{RAED05b}, to construct a subquantum model for both the EPRB and EEPRB experiment.}
This approach has proven fruitful for constructing subquantum models
for many fundamental quantum-physics experiments with photons and neutrons~\cite{MICH14a}.

\section{Discrete-event simulation: general aspects}\label{AIM}

DES is a general methodology for modeling the time evolution of
a system as a discrete sequence of consecutive events~\cite{BANK96,LEEM06}.
DES is used in many different branches of science, engineering, economics, finance, etc.~\cite{BANK96},
but has only fairly recently been adopted as a methodology to construct subquantum models
for basic, fundamental quantum physics experiments~\cite{RAED05b,MICH14a}.

In the following, whenever we use the term DES, we mean DES modeling applied
to quantum physics problems, not DES in general.
The salient features of this particular application of DES are the following.
\begin{itemize}
\item
Events are the basic building blocks of any DES model,
just as points are the basic building blocks of Euclidean geometry.
In DES an event is a {\bf defined} concept,
represented by a model variable taking a particular value (e.g.~a bit changing from zero into one)
at a specific point in time.
In contrast to quantum theory, there is no need to invoke the elusive wave function collapse to ``explain''
how quantum theory may eventually be reconciled with the fact that a measurement yields
a definite yes/no answer, or to appeal to Born's rule. 
\item
It may be difficult to analyze the behavior of a DES model
by means of differential equations, probability theory,
or other mathematical techniques of theoretical physics.
Of course, we may use e.g.~probability or quantum theory
to model the statistics of the data produced by a DES.
\item
It is not practicable to perform a DES without using a digital computer.
A digital computer itself is a physical device that changes
its internal state (all the bits of the CPU and memory) in a discrete,
step-by-step (clock cycle) manner.
Therefore, a DES algorithm running on a digital computer
(which we assume is error-corrected and operating flawlessly)
can be viewed as a metaphor for an idealized experiment
on a physical device (the digital computer)~\cite{RAED16c}.
All aspects of such an experiment are under the control of a programmer.
In the context of EPRB experiments, this means that
any loophole~\cite{LARS14} can be opened or closed at the discretion of the programmer.
For instance, the so-called contextuality loophole, which
is impossible to avoid in a laboratory EPRB experiment~\cite{NIEU17}, can be trivially closed in a DES (see below).
\item
The outcomes of genuine laboratory experiments are subject to unknown influences but
in a DES on a digital computer (operating flawlessly), there are no such influences.
If there were, it would not be possible to {\sl exactly} reproduce the results of a DES.
Therefore, DES is ``the experiment'' to confront a theory with facts
obtained under the same premises as those on which the theory is based.
\item
Although a DES algorithm changes the state of a physical device (the digital computer),
the events and variables in a DES are only metaphors for the ``real'' detector click, etc.
On the other hand, once it has been established that a DES of a subquantum model
yields the correct results, one could build a macroscopic mechanical device that
performs exactly the same as DES.
\item
DES on a digital computer complies with the notion of realism, meaning
that at any time during the DES, the internal state of the digital computer is known exactly,
all variables of the simulation model taking definite values.
Of course, we can always ``hide'' an algorithm and data on purpose.
For instance, we can do this to create the illusion that the ``visible'' data is unpredictable
(a standard technique to generate pseudo-random numbers).
\item
In a digital computer, there are no signals traveling faster than light.
Therefore, on the most basic level, the internal operation of
a digital computer complies with Einstein's notion of local causality.
However, there is nothing that prevents us from performing an
acausal analysis of the data.
For instance, if we generate and store a sequence of numbers and then wish to compute
the sum, we may do this by summing the numbers
in the reverse order of how they were generated.
This trivial example shows that one has to
distinguish between the generation of the raw data and the processing of this data.
For the purpose of constructing a local realist DES model,
it is essential that the process that generates the raw data
complies with Einstein's notion of local causality.
It is not permitted to accumulate data,
perform e.g.~a discrete Fourier transform or compute acausal correlations, and use the results
to describe a quantum physics experiment.
While both these techniques are very useful for a wide variety of data processing tasks~\cite{PRES03},
they are ``forbidden'' in a DES of a subquantum model.
\item
Consistency of the DES methodology demands that a subquantum model for, say, a beam splitter,
must be re-used, without modification, for all experiments in which this beam splitter is used.
Our DES approach seems to satisfy this requirement of consistency,
at least for a vast number of fundamental quantum-physics
experiments with photons and neutrons~\cite{MICH14a}.
Our motivation for considering both the EPRB and extended EPRB (EEPRB) experiments
is to scrutinize the consistency of the DES approach for this category of experiments.
\end{itemize}
Finally, to head off misunderstandings, the DES models that we construct do not, in any way, make use
of the quantum-theoretical predictions for the statistics of the data.
Instead, a DES builds up these statistics by an event-by-event, cause-and-effect, Einstein-local process.
Under appropriate conditions, these statistics can be described by quantum theory, while in other cases they cannot (see below).

\begin{figure}[th]
\centering
\includegraphics[width=0.95\hsize]{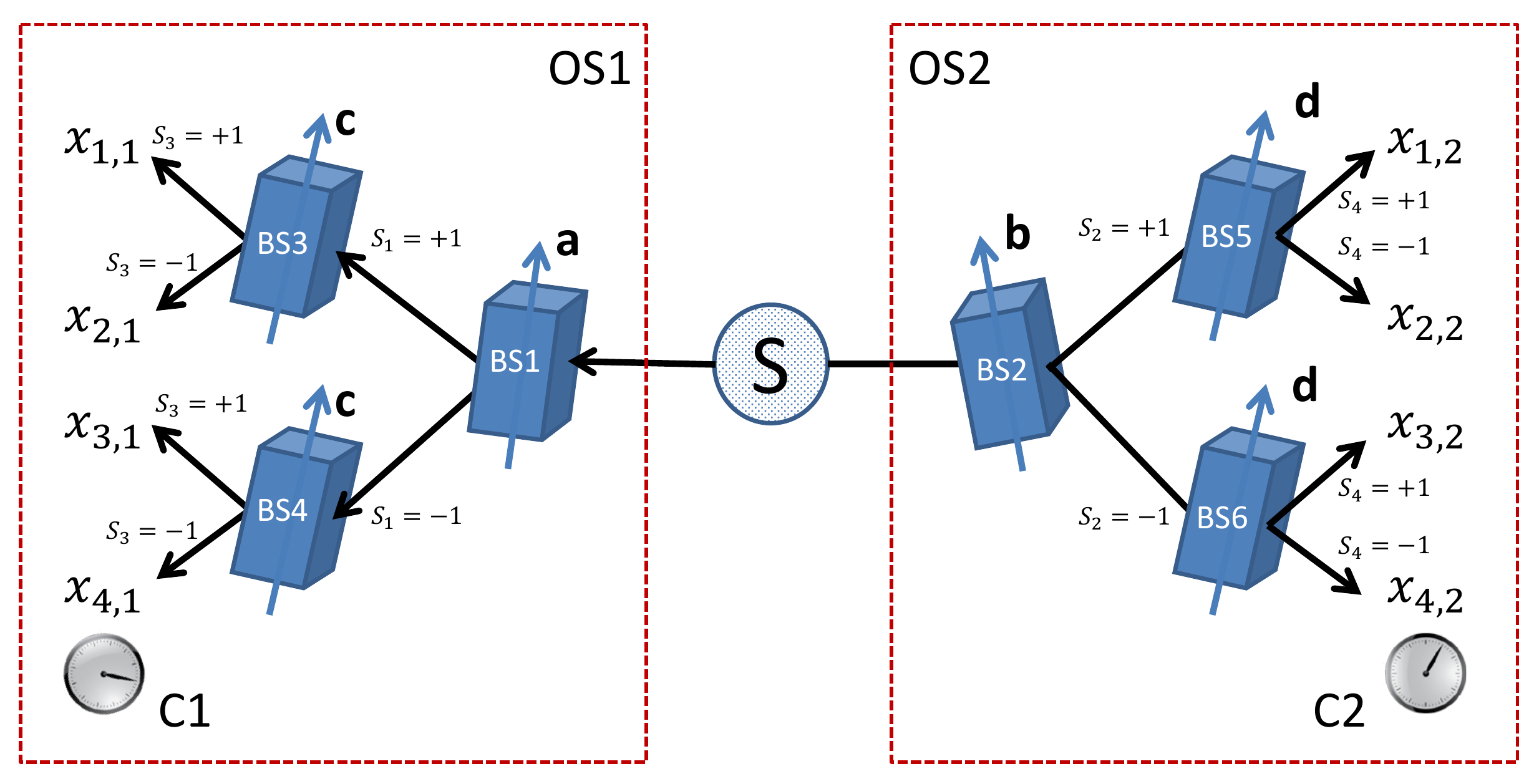}
\caption{
Layout of the extended Einstein-Podolsky-Rosen-Bohm experiment with spin-1/2 particles emitted by the source S.
The observation station OS1 (OS2) contains three beam splitters
BS1, BS3, and BS4, (BS2, BS5, and BS6) and a clock C1 (C2).
The directions of the beam splitters BS1, BS2, BS3, BS4, BS5, and BS6
are represented by vectors $\ba$, $\bb$, $\bc$, $\bc$, $\bd$, and $\bd$, respectively.
}
\label{fig1}
\end{figure}

\section{Extended Einstein-Podolsky-Rosen-Bohm experiment: theory}\label{EEPRB}

\subsection{Thought experiment}\label{TEXP}

Figure~\ref{fig1} shows the layout of an
extended Einstein-Podolsky-Rosen-Bohm (EEPRB) experiment with spin-1/2 particles~\cite{SICA99}.
A source is emitting a pair of particles in two spatially separated directions towards beam splitters BS1 and BS2.
In this idealized experiment, all beam splitters are assumed to be identical,
performing selective (filtering) measurements~\cite{SCHW59,BALL03}.
Selective measurements allow us to attach an attribute with definite value
(e.g. the direction of the magnetic moment or of the polarization) to the particle.
For instance, assuming that BS1, BS3 and BS4 perform ideal selective measurements,
a particle leaving BS1 along path $S_1=+1$ ($S_1=-1$)
will always leave BS3 (BS4) along path $S_3=+1$ ($S_3=-1$) if $\bc=\ba$.
In case of selective measurements, we can attach an attribute to the particle, the value of this attribute being given by $S_1$.
We use the same procedure for attaching attributes to particles leaving the other beam splitters.

Particles leaving BS3,...,BS6 are registered by detectors.
All detectors are assumed to be identical and to have a 100\% detection efficiency (we relax this assumption later).
The binary variables $x_{i,j}=0,1$ for $i=1,\ldots,4$ and $j=1,2$ (see Fig.~\ref{fig1})
indicate which of the four detectors at the left ($j=1$) and right ($j=2$) fire.
For each pair of emitted particles,
exactly one of the detectors on the left
and exactly one of the detectors on the right side of the source will register a particle.
This implies that for $j=1,2$, only one of $x_{1,j}$, $x_{2,j}$, $x_{3,j}$, and $x_{4,j}$ can be nonzero.
The four new variables defined by
\begin{eqnarray}
S_{1}&=&x_{1,1}+x_{2,1}-x_{3,1}-x_{4,1}\;,
\nonumber \\
S_{2}&=&x_{1,2}+x_{2,2}-x_{3,2}-x_{4,2}\;,
\nonumber \\
S_{3}&=&x_{1,1}-x_{2,1}+x_{3,1}-x_{4,1}\;,
\nonumber \\
S_{4}&=&x_{1,2}-x_{2,2}+x_{3,2}-x_{4,2}\;,
\label{EEPRB0}
\end{eqnarray}
all take values $+1$ or $-1$ (see Fig.~\ref{fig1}).
Clearly, $S_1$ and $S_3$ ($S_2$ and $S_4$) encode, in a unique manner,
the path that the left (right) going particle took.
In the following, we use the $S$'s to formulate the DES model.


For practical reasons, most laboratory EPRB experiments are carried out
with photons~\cite{KOCH67,CLAU78,ASPE82b,WEIH98,CHRI13,HENS15,GIUS15,SHAL15},
the polarization of the photons playing the role of the spin.
From a quantum-theoretical viewpoint, there is no loss of generality in doing so because
mathematically, the description of the photon polarization is in terms of Pauli-spin matrix algebra.
In the following, to keep the discussion concise and concrete, we only focus on (E)EPRB experiments
that employ the photon polarization as the ``quantum system'' of interest.

A source is emitting a pair of photons in two spatially separated directions towards beam splitters BS1 and BS2.
BS1 sends one photon of the pair to either BS3 or BS4.
BS2 sends the other photon of the pair to either BS5 or BS6.
In an ideal model, all beam splitters are identical.
Each beam splitter represents a combination of wave plates, an electro-optical modulator (EOM),
and a polarizing beam splitter (PBS), see for example Fig.~1(c) in Ref.~\onlinecite{GIUS15}.
The EOM acts as a switchable (voltage controlled) polarization rotator, the rotation
being characterized by a two-dimensional unit vector (indicated by the arrow through the beam splitter),
relative to local frames of reference attached to the observation stations OS1 and OS2, respectively.

It is expedient to introduce the vectors
$\ba=(\cos a,\sin a)$, $\ba_\bot=(-\sin a,\cos a)$, $\bb=(\cos b,\sin b)$, $\bb_\bot=(-\sin b,\cos b)$,
$\bc=(\cos c,\sin c)$, $\bc_\bot=(-\sin c,\cos c)$, $\bd=(\cos d,\sin d)$, and $\bd_\bot=(-\sin d,\cos d)$.
Photons leave BS1 with linear polarization along either $\ba$ ($S_1=+1$) or  $\ba_\bot$ ($S_1=-1$).
For the other beam splitters, we have similar relations between the direction of the
linear polarization of the photons that leave the beam splitters and the value of the corresponding $S$-variable.


\subsection{Classical Electrodynamics}\label{MAXW}
It is instructive to first consider the case in which the detector signal is linearly proportional
to the intensity of the impinging light.
This case is covered by classical optics, described by Maxwell's theory of electrodynamics.

According to empirical evidence, the intensity of light passing through a polarizer is
given by Malus' law $I=I_0\cos^2(\phi-\psi)$, where $\phi$ is the polarization of the light beam
and $\psi$ is the rotation of the polarizer, both relative to a laboratory frame of reference.
$I_0$ is the intensity of the incident light.

We assume that the source emits ``special'' randomly polarized light towards BS1 and BS2,
special in the sense that the difference between the polarizations of the two beams $\phi_0$ is fixed in time.
Then, using Malus' law for BS1 and BS2, the correlated intensity for one particular, random realization
of the polarization angle $\phi$ is given by
\begin{eqnarray}
I_1(S_1,S_2|\ba,\bb,\phi,\phi_0)&=&I_0^2\frac{1+S_1 \cos2(\phi-a)}{2}\frac{1+S_2 \cos2(\phi-b+\phi_0)}{2}
\;,
\label{MAXW0}
\end{eqnarray}
where $I_0$ denotes the light intensity of a single beam.
Integrating over all polarizations $\phi$ with a uniform density $1/2\pi$ yields
\begin{eqnarray}
I_1(S_1,S_2|\ba,\bb,\phi_0)&=&\frac{I_0^2}{4}\big[1+\frac{1}{2}S_1S_2 \cos2(a-b+\phi_0)\big]
\;.
\label{MAXW1}
\end{eqnarray}
Repeated use of Malus' law and exploiting the fact that the polarizations of the two beams leaving a
beam splitter are orthogonal,
the correlated intensity of the different beams is then given by
\begin{eqnarray}
I(S_1,S_2,S_3,S_4|\ba,\bb,\bc,\bd,\phi_0)=\frac{I_0^2}{16}
\big[1+\frac{1}{2}S_1S_2 \cos2(a-b+\phi_0)\big]\big[1+S_1S_3 \cos2(a-c)\big]\big[1+S_2S_4\cos2(b-d)\big]
\;.
\label{MAXW2}
\end{eqnarray}

The moments of the correlated intensity Eq.~(\ref{MAXW2}) are
\begin{eqnarray}
{\widehat K}_{i}&=&\sum_{S_1,S_2,S_3,S_4=\pm1}S_i I(S_1,S_2,S_3,S_4|\ba,\bb,\bc,\bd,\phi_0)=0\quad,\quad i=1,2,3,4\;,
\nonumber\\
{\widehat K}_{ij}&=&\sum_{S_1,S_2,S_3,S_4=\pm1}S_iS_j I(S_1,S_2,S_3,S_4|\ba,\bb,\bc,\bd,\phi_0)\quad,\quad 1\le i<j \le 4\;,
\nonumber\\
{\widehat K}_{ijk}&=&\sum_{S_1,S_2,S_3,S_4=\pm1}S_iS_jS_k I(S_1,S_2,S_3,S_4|\ba,\bb,\bc,\bd,\phi_0)=0 \quad,\quad i\not=j\not=k\not=i\;,
\nonumber\\
{\widehat K}_{1234}&=&\sum_{S_1,S_2,S_3,S_4=\pm1}S_1S_2S_3S_4I(S_1,S_2,S_3,S_4|\ba,\bb,\bc,\bd,\phi_0)=
I_0^2\cos2(a-c)\cos2(b-d)
\;.
\label{MAXW3}
\end{eqnarray}
For $\phi_0=\pi/2$ (orthogonally polarized beams) and $I_0=1$, the explicit expressions for the two-$S$ correlations are
\begin{eqnarray}
\begin{array}{lclclc}
{\widehat K}_{12}=-\frac{1}{2}\cos2(a-b)&,&{\widehat K}_{13}=\cos2(a-c)&,&
{\widehat K}_{14}=-\frac{1}{2}\cos2(a-b)\cos2(b-d)\quad,& \\ \\
{\widehat K}_{23}=-\frac{1}{2}\cos2(a-b)\cos2(a-c)&,&{\widehat K}_{24}=\cos2(b-d)&,&
{\widehat K}_{34}=-\frac{1}{2}\cos2(a-b)\cos2(a-c)\cos2(b-d)&. \\
\end{array}
\label{MAXW4}
\end{eqnarray}
The factor 1/2 which appears in Eq.~(\ref{MAXW1}) and in four of the six second moments ${\widehat K}_{ij}$
is characteristic of the correlation of two light intensities.
Here and in the following, we use the hat on top of the symbols to emphasize that the expressions
have been obtained from a theoretical model.

\subsection{Quantum theory}\label{SQT}

Classical electrodynamics describes the intensity of light and does not discriminate between individual events.
In contrast, quantum theory can be used to describe the statistics of events, particularly in cases, such as the EPRB experiment, where detectors can discriminate between them.

For a pair of photons whose polarizations are described by the singlet state,
Appendix~\ref{QT} shows that the joint probability to observe one photon in the path labeled by $S_1$ and
the other one in the path labeled by $S_2$ is given by~\cite{RAED07c}
\begin{eqnarray}
P(S_1,S_2|\ba,\bb,Z)=\frac{1-S_1S_2\;\cos2(a-b)}{4}
,
\label{SQT0}
\end{eqnarray}
where $Z$ denotes a valid proposition that represents all conditions under which
the experiment is performed with the exception of $\ba$ and $\bb$.
Note that for $\phi_0=\pi/2$, Eq.~(\ref{MAXW1}) differs from Eq.~(\ref{SQT0}) through the factor 1/2 only.

In Appendix~\ref{QT}, we show that
the joint probability to observe one photon in the path labeled by $(S_1,S_3)$ and the other in the path labeled by $(S_2,S_4)$
is given by
\begin{eqnarray}
P(S_1,S_2,S_3,S_4|\ba,\bb,\bc,\bd,Z)=\frac{1}{16}\big[1-S_1S_2\cos 2(a-b)\big]
\big[1+S_1S_3 \cos2(a-c)\big]\big[1+S_2S_4\cos2(b-d)\big]
\;.
\label{SQT1}
\end{eqnarray}
As already mentioned in the introduction, a subquantum model for the EEPRB experiment must not make use of Eq.~(\ref{SQT1}) to generate the quadruples $(S_1,S_2,S_3,S_4)$.

Note that the only nontrivial difference between Eq.~(\ref{MAXW2}) and Eq.~(\ref{SQT1})
is that in the former case, the absolute value of the prefactor of the $S_1S_2$ term never exceeds 1/2
whereas in the latter case, it is equal to minus one.
The expressions of the second moments of Eq.~(\ref{SQT1}) read
\begin{eqnarray}
\begin{array}{lclclc}
{\widehat E}_{12}=-\cos2(a-b)&,&{\widehat E}_{13}=\cos2(a-c)&,&
{\widehat E}_{14}=-\cos2(a-b)\cos2(b-d)\quad,\\ \\
{\widehat E}_{23}=-\cos2(a-b)\cos2(a-c)&,&{\widehat E}_{24}=\cos2(b-d)&,&
{\widehat E}_{34}=-\cos2(a-b)\cos2(a-c)\cos2(b-d)&,\\
\end{array}
\label{SQT2}
\end{eqnarray}
which are not all equal to the corresponding expressions of the ${\widehat K}$'s, see Eq.~(\ref{MAXW4}).
From Eq.~(\ref{SQT1}) it follows that
${\widehat E}_{1}={\widehat E}_{2}={\widehat E}_{3}={\widehat E}_{4}={\widehat E}_{123}={\widehat E}_{124}={\widehat E}_{134}={\widehat E}_{234}=0$ and
${\widehat E}_{1234}=\cos2(a-c)\cos2(b-d)$.

Clearly, {\color{black}in order to have a subquantum model generate data that agrees either with Maxwell's theory Eq.~(\ref{MAXW2}) or quantum theory Eq.~(\ref{SQT1}),}
we only have to construct a subquantum model in which
we can control the prefactor of the $S_1S_2$ term in Eq.~(\ref{MAXW2}).
Thinking of light as a collection of photons,
in sections~\ref{PRPI} and~\ref{SUBM}, we explain how this control naturally results from the simple fact
that we have to classify individual events as photons
or something else, whereas in the ``classical'' case this classification is not an issue.
{\color{black}
At this point, it should be mentioned that within the context of  the classical and quantum theory of light, changing the prefactor $1/2$ of the $S_1S_2$ term in Eq.~(\ref{MAXW2}) is a subtle issue, intimately related to the amount of second-order coherence one can observe by measuring either intensities or by counting clicks of a detector~\cite{KHRE20}. A discussion of this important issue is out of the scope of this paper and we refer the reader who is interested in these aspects to the in-depth analysis given in Ref.~\onlinecite{KHRE20}. In our paper, Eqs.~(\ref{MAXW2}) and~(\ref{SQT1})
are only used to provide the classical/quantum results which any valid subquantum model for the (E)EPRB experiment
has to reproduce.
}

An important feature of this EEPRB experiment is that all the correlations that
are required to test for violations of Bell/Clauser-Horne-Shimony-Holt (CHSH) inequalities~\cite{CLAU69,BELL93}
are obtained in a single run (instead of three/four runs) of the experiment~\cite{SICA99}.
The EEPRB experiment does not suffer from the contextuality loophole~\cite{NIEU17}.
As $0\le P(S_1,S_2,S_3,S_4|\ba,\bb,\bc,\bd,Z)\le 1$, it follows directly
that all Bell-type inequalities, including all variants of the CHSH inequality,
can never be violated~\cite{RAED11a}. This is easily seen by evaluating the sum
$\sum_{S_1,S_2,S_3,S_4=\pm1}g(S_1,S_2,S_3,S_4)P(S_1,S_2,S_3,S_4|\ba,\bb,\bc,\bd,Z)$
for various choices of the function $g(S_1,S_2,S_3,S_4)$ such as
$-1\le g(S_1,S_2,S_3,S_4)=S_1S_2+S_1S_3+S_2S_3\le3$, or $-2\le g(S_1,S_2,S_3,S_4)=S_1S_3+S_1S_4+S_2S_3-S_2S_4\le2$,
for example.
In words, the quantum-theoretical description of the EEPRB experiment predicts that all Bell/CHSH inequalities are satisfied,
in stark contrast to the case in which the correlations that enter the Bell/CHSH inequalities are computed from
the quantum-theoretical description of the EPRB experiment.

\subsection{Practical realization: photon identification problem}\label{PRPI}

The exposition in subsection~\ref{TEXP} assumes that
each emitted pair of particles triggers exactly two detectors,
namely only one of the four detectors at OS1
and one of the four detectors at OS2.
In a laboratory experiment with Stern-Gerlach magnets and magnetic billiard balls,
this assumption may hold true.
However, it is not at all evident to have a source which
only creates correlated pairs of elementary particles such as photons
which upon hitting a detector, will trigger exactly one detector at OS1
and exactly one detector at OS2.

With the exception of two experiments~\cite{GIUS15,SHAL15}, EPRB experiments with photons
use time coincidence to identify photon pairs~\cite{KOCH67,CLAU78,ASPE82b,WEIH98,CHRI13,HENS15}.
The two EPRB experiments~\cite{GIUS15,SHAL15} that do not rely on time coincidence
employ local, adjustable voltage thresholds to identify photons.
This procedure is mathematically equivalent
to attaching a local time tag to each particle, or to using time coincidence~\cite{RAED17a}.
Therefore, in the following, we only discuss the subquantum model that uses local time tags as the vehicle
for identifying pairs of photons. The modifications required to deal with voltage thresholds
are trivial~\cite{RAED17a}.

As explained above, a minimal theoretical model of a laboratory (E)EPRB experiment
with photons should include a procedure to identify (pairs of) photons.
Specifically, not including the data by which the photons and/or pairs
are identified opens the so-called photon identification loophole~\cite{RAED17a}.
By design, the EPRB experiments
that claim to be loophole free~\cite{HENS15,GIUS15,SHAL15} all suffer from this loophole.

As shown in Fig.~\ref{fig1}, observation stations OS1 and OS2 are equipped with local clocks C1 and C2, respectively.
The time $t_1$ ($t_2$) at which a detector in OS1 (OS2) fires is read off from the local clock C1 (C2).
The clocks C1 and C2 are synchronized before the source starts to emit pairs of particles
and, being ideal clocks, remain synchronized for the duration of the experiment.
Similarly, the frames of reference of OS1 and OS2 are aligned before the
source starts to emit pairs of particles and do not change afterwards.

Concretizing the aim of this paper, in the next section, we describe a local realist
model of the (E)EPRB experiment that reproduces the statistical predictions
of quantum theory given by Eqs.~(\ref{SQT0}) and~(\ref{SQT1}).
In formulating the DES model, we call the agents that carry the information
from the source to the observation stations ``photons'' and use the language of optics.

\section{Subquantum model}\label{SUBM}

Before describing all the components of the subquantum model, we
recall the basic strategy that we adopt in constructing such a model.
As quantum theory describes the most ideal version of the (E)EPRB experiment
and as our aim is to show that the subquantum model reproduces the results of the former,
we construct a DES of the most ideal version of the (E)EPRB experiment.
The DES model for the EEPRB experiment that we describe next contains
the ideal implementation of EPRB laboratory experiments.

In concert with our general strategy to set up the subquantum model,
we assume that the source emits pairs of photons only and this at regular time intervals $\Delta$.
The time at which the $n$th pair is emitted is given by $T_n=n\Delta$.
The time it takes for a photon to travel from the source to BS1 or BS2 is assumed
to be constant and the same for all photons traveling to OS1 and OS2.
We denote this time of flight by $T'_{\mathrm{TOF}}$.
Similarly, the time of flight from BS1 to BS3 or BS4 (BS2 to BS5 or BS6) and the time of flight from
BS3, BS4, BS5 or BS6 to the corresponding detectors
are denoted by $T''_{\mathrm{TOF}}$ and $T'''_{\mathrm{TOF}}$, respectively.
The total time of flight is then $T_{\mathrm{TOF}}=T'_{\mathrm{TOF}}+T''_{\mathrm{TOF}}+T'''_{\mathrm{TOF}}$.

In the DES model, the polarization of the photon traveling to BS1 (BS2)
is represented by a two-dimensional unit vector $\bx_1=(\cos\phi,\sin\phi)^{\mathrm{T}}$
($\bx_2=(-\sin\phi,\cos\phi)^{\mathrm{T}}$).
The angle $\phi$ is chosen to be uniformly random from the interval $[0,2\pi)$.
As $\bx_1^{\mathrm{T}}\cdot\bx_2=0$, the polarizations of the photons of each pair are orthogonal,
that is, they are maximally anticorrelated and randomly distributed over the unit circle.

In the following, we specify the DES rules for BS1 only.
The rules for the other beam splitters are identical and are obtained by a simple change of symbols.
In the DES model, the operation of beam splitter BS1 is defined by the rules
\begin{eqnarray}
S_1&=&
\left\{
\begin{array}{lcr}
+1 & \mathrm{if} & \cos^2 (\phi-a)> r \\
-1 & \mathrm{if} & \cos^2 (\phi-a)\le r\\
\end{array}
\right.
\quad,\quad
\bx_1^\prime=\left\{
\begin{array}{lcr}
\ba&\mathrm{if}& S_1=+1\\
\ba_\bot&\mathrm{if}& S_1=-1\\
\end{array}
\right.
\;,
\label{SUBM0}
\end{eqnarray}
where $0<r<1$ denotes a uniform pseudo-random number.
Here and in the following, it is implicitly understood that
a new instance of the pseudo-random number $r$ is generated
with each invocation of an equation in which $r$ appears.
The unit vector $\bx_1^\prime$ denotes the polarization of the photon leaving BS1.
It is not difficult to see that the model defined by Eq.~(\ref{SUBM0})
generates $S_1=+1$ ($S_1=-1$) events with a relative frequency given by
$\cos^2 (\phi-a)$ ($\sin^2 (\phi-a)$), that is, Eq.~(\ref{SUBM0}) produces data that is in concert with Malus' law
if the polarization of the incident photon is constant in time.

Optical components such as wave plates and EOMs contain birefringent material
which changes the polarization by retarding (or delaying) one component of the
polarization with respect to its orthogonal component.
In the DES, this retardation effect is accounted for
by assuming that as a photon passes through a beam splitter, it may suffer from a time delay
which may depend on the direction of the beam splitter relative to the polarization of the photon.

Obviously, the law of retardation in the subquantum model cannot be derived
from Maxwell's theory or quantum theory.
We can only find the subquantum law of retardation by trial and error.
Fortunately, from earlier work we already know the subquantum law of retardation
for the EPRB experiment~\cite{RAED06c,ZHAO08} and we only need to extend this law slightly
to have the DES reproduce the quantum-theoretical results for both the EPRB and EEPRB experiment.
Specifically, for BS1, the two DES rules for the subquantum law of retardation read
\begin{eqnarray}
\tau_1&=&
\tau_{\mathrm{EPRB}}(\bx_1,\ba) \left|\frac{1-\bx_1\cdot{\bu}_1}{2}\right|^{\beta}
=r^\prime T_{\mathrm{max}}\left|\sin 2(\phi-a)\right|^\alpha\left|\frac{1-{\bx}_1\cdot{\bu}_1}{2}\right|^{\beta}
\quad,\quad{\bu}_1 \leftarrow\bx_1
\;,
\label{SUBM1aa}
\end{eqnarray}
where $\bx_1$ (or equivalently $\phi$) is the polarization of the incoming photon,
$0<r^\prime<1$ is another uniform pseudo-random number,
$T_{\mathrm{max}}$ is an adjustable parameter specifying maximum retardation,
and $\alpha>0$ is an adjustable parameter controlling the dependence
of the retardation on the difference between the photon polarization $\bx_1$
and the orientation of the beam splitter $\ba$.
As indicated by the subscript EPRB, $\tau_{\mathrm{EPRB}}=r^\prime T_{\mathrm{max}}\left|\sin 2(\phi-a)\right|^\alpha$ suffices
to reproduce the quantum-theoretical results of the EPRB experiment~\cite{RAED06c,RAED07c,ZHAO08,RAED16c,RAED17a}.

The new features are the last factor in Eq.~(\ref{SUBM1aa}), $\beta>0$ being an adjustable parameter,
and the rule ${\bu}_1 \leftarrow\bx_1$ which updates the two-dimensional vector $\bu_1$.
The initial value of $\bu_1$ can be any vector that has a norm less than or equal to one.
This vector is attached to the beam splitter and may be thought of
as representing (on a subquantum level) the electrical polarization of the material~\cite{MICH14a}.

The purpose of the factor $|(1-{\bx}_1\cdot{\bu}_1)/2|^{\beta}$ in Eq.~(\ref{SUBM1aa})
is to turn off the generation of random retardation times if the polarization of the incoming photons is constant.
To see how this works, first consider the case that the polarization of the incoming photons is constant,
say $\bx_1={\widetilde\bx}_1$.
Then, after the first photon has passed by,
$\bu_1={\widetilde\bx}_1$ and $|(1-{\bx}_1\cdot{\bu}_1)/2|^{\beta}=0$ for all photons that follow.
Next, assume that the polarization of the photons entering BS1 is randomly
distributed over the unit circle. Then, because ${\bx}_1$ and ${\bu}_1$ (which is equal to the
polarization ${\bx}_1$ of the previous photon) are independent, $|(1-{\bx}_1\cdot{\bu}_1)/2|^{\beta}$ 
is just a random variable in $[0,1]$ multiplying $\tau_{\mathrm{EPRB}}$.
The idea to store and use the value of the polarization of the previous
photon has also been used to reproduce, by DES, the quantum-theoretical results for a
large variety of single-photon and single-neutron experiments~\cite{RAED05b,MICH14a}.
The capability of the subquantum model to disable the generation of random retardation times
if the polarization of the incoming photons is constant
is essential for reproducing the quantum-theoretical results of both the EPRB and EEPRB
experiments with the same subquantum model.

At this point, it may be of interest to mention that on the basis of the statistics
(i.e.~averages and correlations) only, it is not possible to make statements
about the uniqueness of the subquantum law of retardation.
As a matter of fact, in the case at hand, replacing Eq.~(\ref{SUBM1aa}) by the rules
\begin{eqnarray}
\tau_1&=&
\tau_{\mathrm{EPRB}}(\bx_1,\ba) \left|\frac{1-{\bu}_1\cdot{\bu}_1}{2}\right|^{\beta}
=r^\prime T_{\mathrm{max}}\left|\sin 2(\phi-a)\right|^\alpha\left|\frac{1-{\bu}_1\cdot{\bu}_1}{2}\right|^{\beta}
\;,
\label{SUBM1a}
\\
{\bu}_1 &\leftarrow&\gamma {\bu}_1 + (1-\gamma)\bx_1
\;,
\label{SUBM1b}
\end{eqnarray}
works equally well.

Equation~(\ref{SUBM1b}) defines a deterministic learning machine (DLM)~\cite{RAED05b,MICH14a},
which learns, event-by-event, the time average of the polarizations $\bx_1$ carried by the photons.
The speed and accuracy by which $\bu_1$ approaches the time average of the $\bx_1$'s
is controlled by the parameter $0< \gamma<1$~\cite{MICH14a}.
The order in which Eqs.~(\ref{SUBM1a}) and~(\ref{SUBM1b}) are executed is irrelevant.
The DLM defined by Eq.~(\ref{SUBM1b}) is the same as the one that
has been used to reproduce, by DES, the quantum-theoretical results for a
large variety of single-photon and single-neutron experiments~\cite{RAED05b,MICH14a}.
If the polarizations of the incoming photons are constant, say ${\widetilde\bx}_1$, and
a certain number (depending on $\gamma$) of photons has passed by, we have
$\bu_1\approx{\widetilde\bx}_1$ and $|(1-{\bu}_1\cdot{\bu}_1)/2|^{\beta}\approx0$.
If the polarizations of the photons entering BS1 are randomly
distributed over the unit circle, $\bu_1\to0$ and $|(1-{\bu}_1\cdot{\bu}_1)/2|^{\beta}\approx1$.
Then, just as in the case of Eq.~(\ref{SUBM1aa}),
the factor $|(1-{\bu}_1\cdot{\bu}_1)/2|^{\beta}$ in Eq.~(\ref{SUBM1a}) is used
to turn off the generation of random retardation times if the polarizations of the incoming photons are constant.

In our idealized experiment, all detectors are assumed to be identical and to have a 100\% detection efficiency.
After a photon has passed BS3 or BS4 (BS5 or BS6), it may trigger one and only one detector in OS1 (OS2),
symbolized by one of $x_{1,1}$, $x_{2,1}$, $x_{3,1}$, or $x_{4,1}$
($x_{1,2}$, $x_{2,2}$, $x_{3,2}$, or $x_{4,2}$)
being equal to one and the other ones being equal to zero.
The time $t_1$ ($t_2$) at which a detector in OS1 (OS2) fires is read off from the local clock C1 (C2).
These local clocks C1 and C2 are synchronized before the source starts to emit pairs of photons
and, being ideal clocks, remain synchronized for the duration of the experiment.
For the $n$th emitted pair, the arrival times are given by
\begin{eqnarray}
t_{1,n}&=&T_{\mathrm{TOF}}+n\Delta+\tau_{1,n} +
\left\{
\begin{array}{lcl}
\tau_{3,n}&\mathrm{if}&S_1=+1 \\ \\
\tau_{4,n}&\mathrm{if}&S_1=-1 \\
\end{array}
\right.
\\ \nonumber \\
t_{2,n}&=&T_{\mathrm{TOF}}+n\Delta+\tau_{2,n}+
\left\{
\begin{array}{lcl}
\tau_{5,n}&\mathrm{if}&S_2=+1 \\ \\
\tau_{6,n}&\mathrm{if}&S_2=-1 \\
\end{array}
\right.
\;,
\label{SUBM3}
\end{eqnarray}
where we have attached the subscript $n$ to keep track of which pair of the emitted pairs
we are dealing with.
For each pair-emission event $n=1,\ldots,N$, Eqs.~(\ref{SUBM0})~--~(\ref{SUBM3})
generate the data $(S_{1,n},S_{3,n},t_{1,n})$ and $(S_{2,n},S_{4,n},t_{2,n})$.
Note that OS1 and OS2 only share the angle of polarization $\phi$ characterizing the pair of photons,
nothing else.

In each triple in $(S_{i,n},S_{i+2,n},t_{i,n})$, we replace the time variable
by a (local) binary variable $w_{i,n}$ to indicate whether a detection event
is classified as a photon ($w_{i,n}=1$) or not ($w_{i,n}=0$).
Specifically, the rule to decide whether a detection event
corresponds to the observation of a photon or of something else is given by
\begin{eqnarray}
w_{i,n}=\left\{\begin{array}{cclcl}
1 & \mathrm{if} & 0\le t_{i,n}-T_{\mathrm{TOF}}-n\Delta \le W&,& \hbox{``a photon''}\\
\\
0 &  \mathrm{if} & t_{i,n}-T_{\mathrm{TOF}}-n\Delta > W &,& \hbox{``something else''}\\
\end{array}\right.
\quad,\quad i=1,2
\;,
\label{SUBM4}
\end{eqnarray}
where $W$ is the time window (an adjustable parameter).
We emphasize that the decision process defined by Eq.~(\ref{SUBM4}) only involves
variables that are local to the observation stations.

Equations~(\ref{SUBM0})~--~(\ref{SUBM4}) define the rules by which the subquantum model generates the
data sets
\begin{eqnarray}
{\cal S}_1=\big\{(S_{1,n},S_{3,n},w_{1,n})\;|\;n=1,\ldots,N\big\}
\quad\mathrm{and}\quad
{\cal S}_2=\big\{(S_{2,n},S_{4,n},w_{2,n})\;|\;n=1,\ldots,N\big\}
\;,
\label{SUBM5}
\end{eqnarray}
collected by OS1 and OS2, respectively.
From the data sets ${\cal S}_1$ and ${\cal S}_2$, we compute the single- and two-particle averages
\begin{eqnarray}
\begin{array}{lclcl}
{K}_i=\frac{1}{N}\sum_{n=1}^N S_{i,n}
&,&
E_i=\frac{\sum_{n=1}^N w_{1,n}w_{2,n}S_{i,n}}{\sum_{n=1}^N w_{1,n}w_{2,n}}
&,&i=1,2,3,4\\
\\
{K}_{ij}=\frac{1}{N}\sum_{n=1}^N S_{i,n}S_{j,n}
&,&
E_{ij}=\frac{\sum_{n=1}^N w_{1,n}w_{2,n}S_{i,n}S_{j,n}}{\sum_{n=1}^N w_{1,n}w_{2,n}}
&,&(i,j)=(1,2),(1,3),(1,4),(2,3),(2,4),(3,4)
\end{array}
\;,
\label{SUBM6}
\end{eqnarray}
without ($K$'s) and with ($E$'s) the photon identification process in place.

\section{Simulation results}\label{SEPRB}

\begin{figure}[ht]
\begin{center}
\includegraphics[width=0.45\hsize]{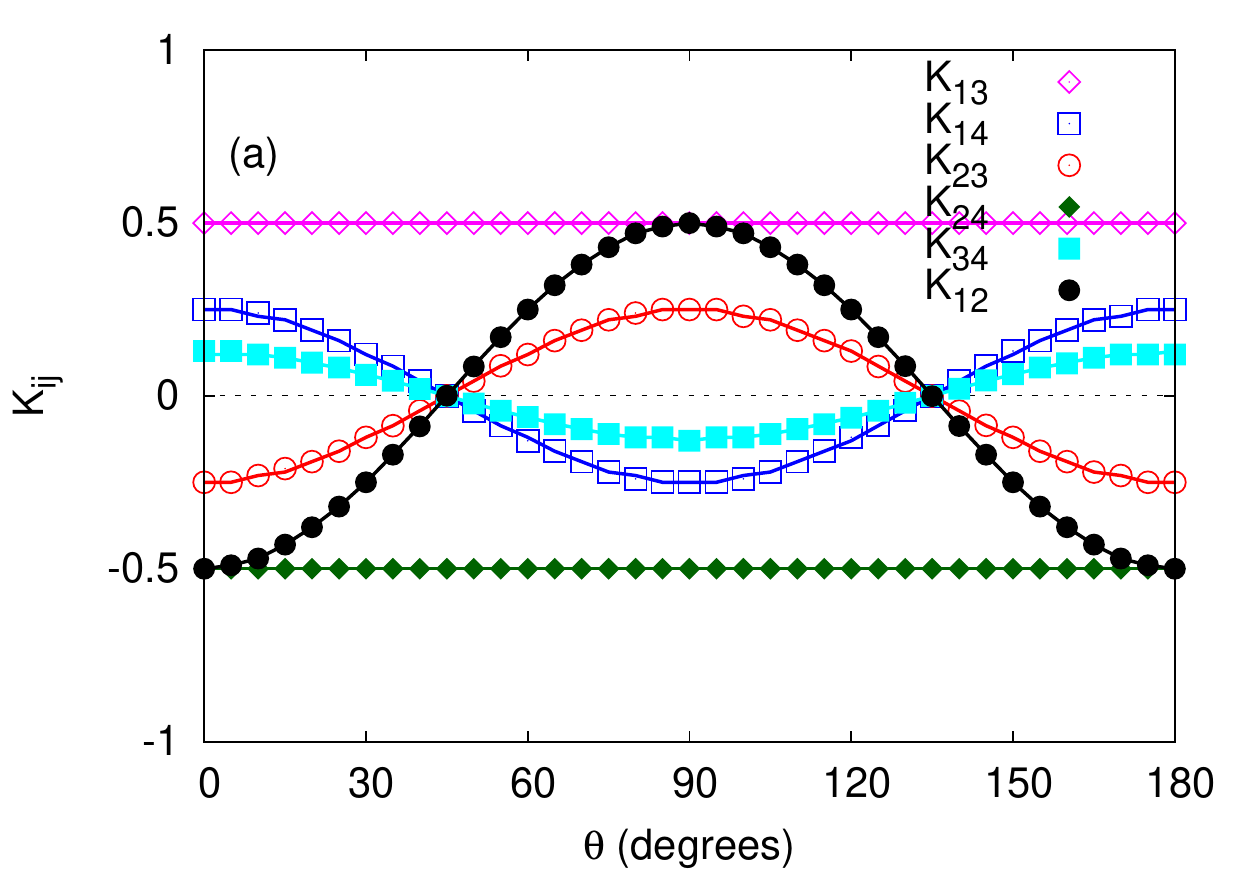}
\includegraphics[width=0.45\hsize]{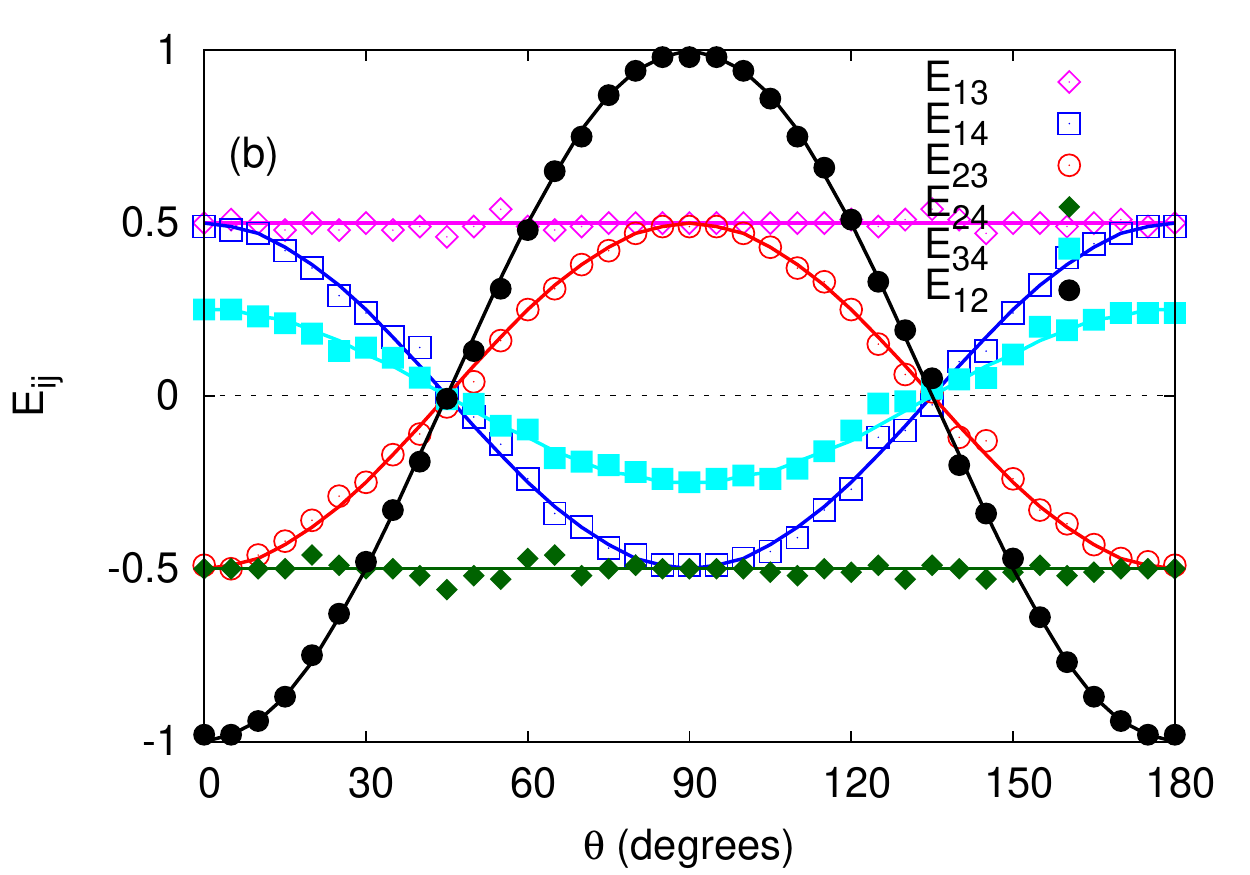}
\end{center}
\caption{DES results for the correlations between all pairs of the $S$-variables
as a function of $\theta=a-b$, with $b=0$, $c=a+\pi/6$ and $d=\pi/3$.
The source emits pairs of photons with orthogonal polarizations chosen randomly.
Ignoring statistical fluctuations, the averages of all $S$-variables (not shown) are zero.
(a) The correlations $K_{ij}$ computed without photon identification (markers) are
in excellent agreement with the corresponding correlations ${\widehat K}_{ij}$ (see Eq.~(\ref{MAXW4}))
predicted by Maxwell's theory (solid lines) for two light beams with orthogonal, random polarization ($\phi_0=\pi/2)$;
(b) The correlations $E_{ij}$ computed with photon identification (markers) are
in excellent agreement with the corresponding correlations ${\widehat E}_{ij}$
predicted by quantum theory (solid lines, see Eq.~(\ref{SQT2}))
for two spin-1/2 particles in a singlet state.
}
\label{fig2}
\end{figure}

\begin{figure}[ht]
\begin{center}
\includegraphics[width=0.45\hsize]{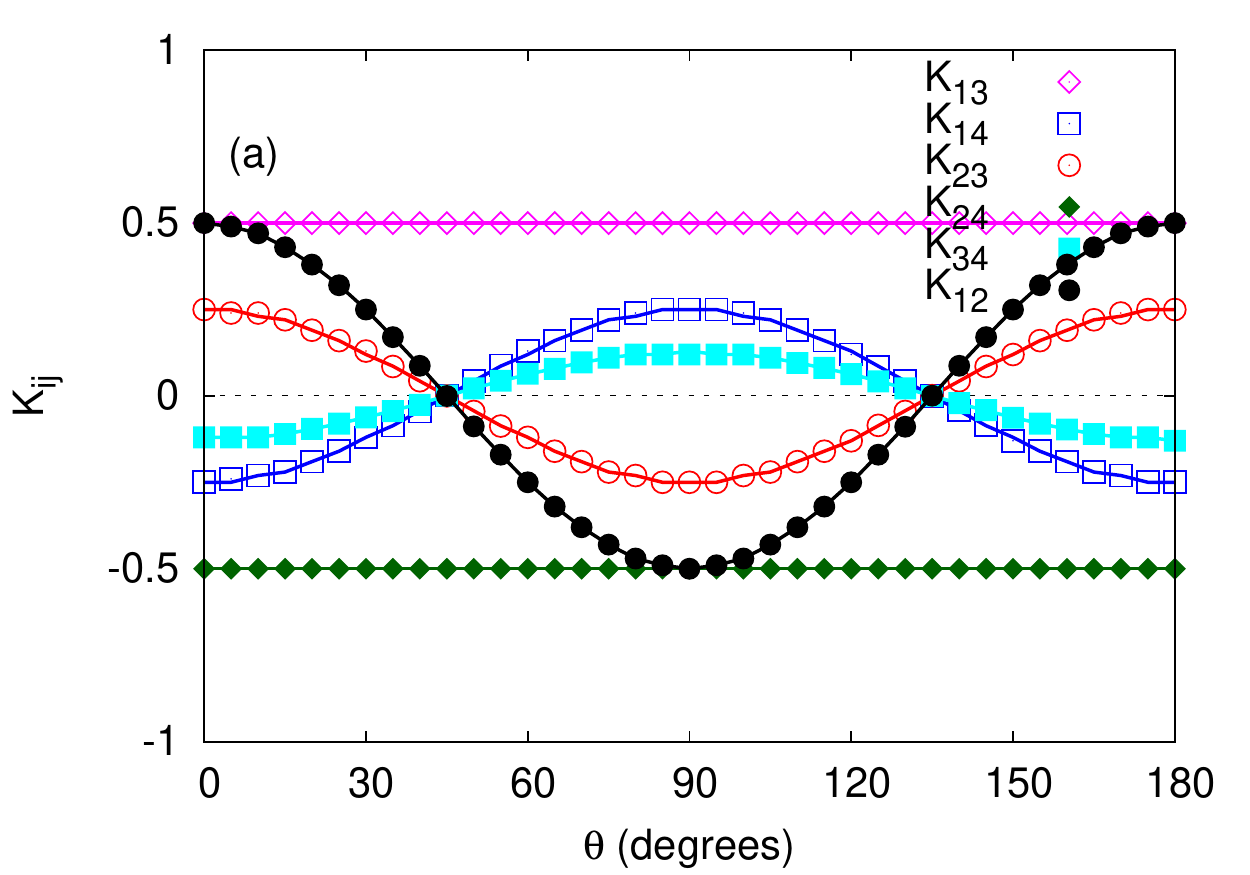}
\includegraphics[width=0.45\hsize]{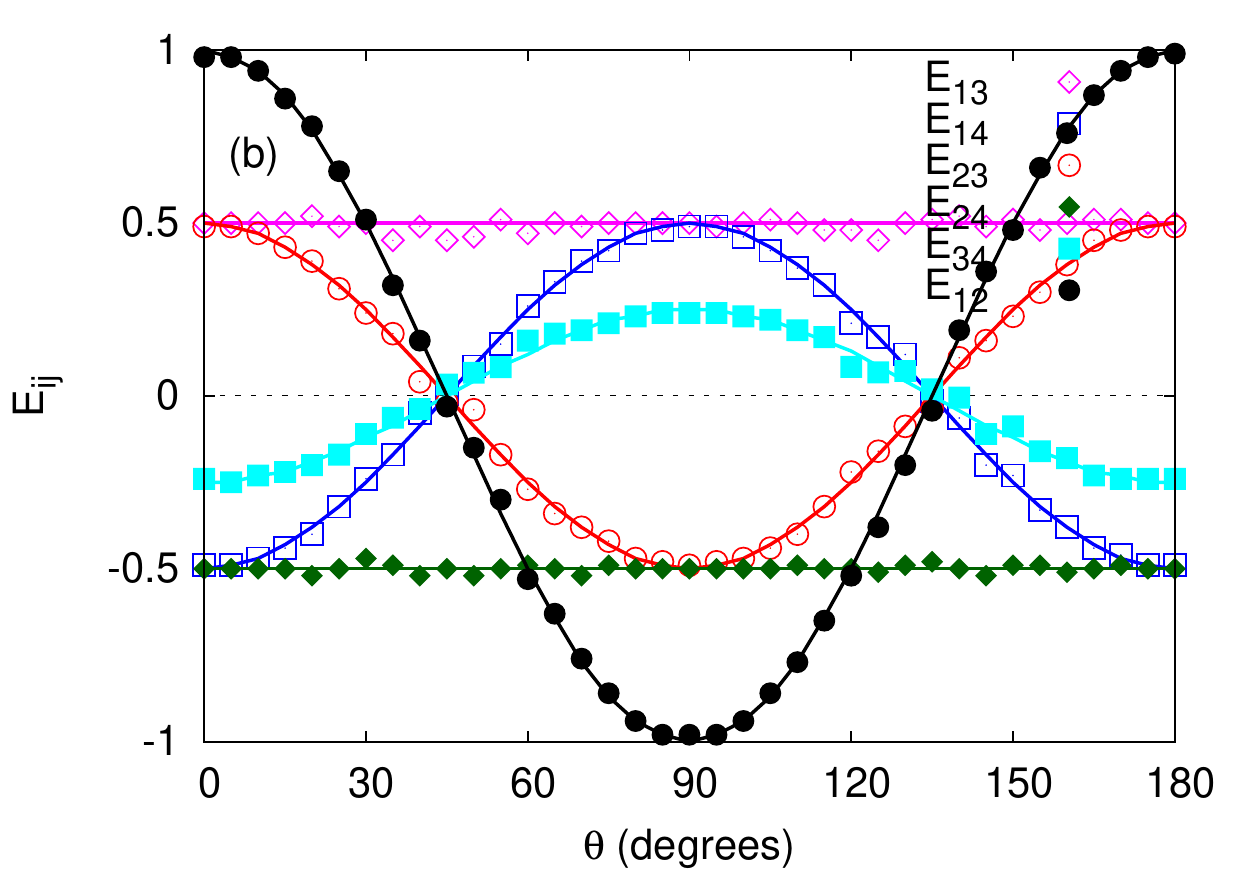}
\end{center}
\caption{Same as Fig.~\ref{fig2}
except that the source emits pairs of photons with the same instead of orthogonal polarizations
chosen randomly.
(a) The correlations $K_{ij}$ computed without photon identification (markers) are
in excellent agreement with the corresponding correlations ${\widehat K}_{ij}$
(solid lines) predicted by Maxwell's theory
for two light beams with the same, random polarization ($\phi_0=0$);
(b) The correlations $E_{ij}$ computed with photon identification (markers)
cannot be obtained from quantum theory for two spin-1/2 particles (see appendix~\ref{OT}) but
are in excellent agreement with the corresponding correlations (solid lines)
obtained from the expression Eq.~(\ref{SQT1}) in which
the factor $\big[1-S_1S_2\cos 2(a-b)\big]$ is replaced by $\big[1+S_1S_2\cos 2(a-b)\big]$.
}
\label{fig3}
\end{figure}

Below, we specify the DES parameters that have been used,
discuss the data shown in Figs.~\ref{fig2} and~\ref{fig3}, and provide additional information
about simulation data that we do not show.

\begin{itemize}
\item
The number of emitted pairs is $N=1000000$ per setting $(\ba,\bb,\bc,\bd)$.
\item
The maximum retardation time was chosen to be $T_{\mathrm{max}}=5000$ (dimensionless units).
Pairs of particles are emitted with a time interval $\Delta>2T_{\mathrm{max}}$.
In line with our strategy to perform an ideal experiment, this choice
eliminates the possibility of misidentifying pairs
and also ensures that at each instant of time, there is only
one photon in transit to OS1 and only one other photon en route to OS2.
For $T_{\mathrm{TOF}}$ we can take any nonnegative value.
In fact, from Eq.~(\ref{SUBM4}) it follows that the
actual values of $\Delta$ and $T_{\mathrm{TOF}}$ do not enter in the DES algorithm.
\item
In Fig.~\ref{fig2}(a), we show the DES results for the case
without photon identification (that is if $W>T_{\mathrm{max}}$ or $\alpha=\beta=0$).
Then the DES reproduces
the results of Maxwell's theory, {\bf by an event-by-event process}~\cite{MICH11a}.
Specifically, if the polarization of the incoming photon is constant,
the DES model of the beam splitter itself generates data according to Malus' law.
The DES model of the EEPRB experiment with randomly polarized light produces data that,
ignoring statistical fluctuations, is in excellent agreement with Eq.~(\ref{MAXW4}),
see Fig.~\ref{fig2}(a).
\item
Using a local time window $W=1$ (dimensionless units) and for $\alpha=4$ and $\beta=1/2$,
the {\bf event-by-event process} yields
results (see Fig.~\ref{fig2}(b)) for the $E$'s which are in excellent
agreement with quantum theory
(data for the first, third and fourth moments are not shown).
The ratio of identified photon pairs to emitted pairs depends on $a$ and $b$ and varies between approximately $11\%$ for $a=b$ and $0.1\%$ for $|a-b|=\pi/4$.
For $W=8$, this ratio changes to
approximately $18\%$ for $a=b$ and $0.8\%$ for $|a-b|=\pi/4$,
while the agreement with quantum theory is still very good (data not shown).
\item
The data obtained by identifying photons using time coincidence instead of local time windows
are almost the same and are therefore not shown.
In the limit $W/T_{\mathrm{max}}\rightarrow0$, $N\rightarrow\infty$, and $\alpha=4$, it has been proven
analytically that the DES model of the EPRB experiment
yields the correlation $E_{12}=-\cos2(a-b)$ exactly~\cite{ZHAO08}.
\item
Using the same pseudo-random sequence for each choice of settings
renders the DES compliant with the notion of a counterfactually definite theory.
In this case, the DES results (data not shown) are, for all practical purposes,
the same as those obtained by using different pseudo-random sequences for each choice of settings.
Operating in this mode, the subquantum model of the EPRB experiment does not suffer from
the contextuality loophole~\cite{NIEU17} nor of any other known loopholes~\cite{LARS14}.
This confirms the conclusion of an earlier work~\cite{RAED16c}, which adopted a different approach
to realize counterfactually-definite compliant simulations.
The demonstration that there exist both counterfactually-definite and non-counterfactually-definite compliant
computer models for the EPRB experiments that produce results in complete
agreement with those of quantum theory implies that,
for the case of EPRB experiments, counterfactual definiteness is not incompatible with quantum physics~\cite{RAED17a}.
\item
In the DES, it is trivial to account for the detection efficiency $0\le\eta\le1$.
For each detection event, we generate a pseudo-random number $r''$
and remove the detection event from the data set if $r''>\eta$.
We find that the only effect of reducing $\eta$ is to increase
the statistical fluctuations (data not shown).
The agreement with quantum theory is not affected.
\item
In Fig.~\ref{fig3}, we show the DES results for the case in
which the source emits photons with the same polarization chosen randomly.
In this case, the DES reproduces the results of Maxwell's theory (Fig.~\ref{fig3}(a)).
A corresponding quantum-theoretical result does not exist, see appendix~\ref{OT}.
However, the DES data are in excellent agreement with a {\bf non-quantum} probabilistic theory
in which the factor $\big[1-S_1S_2\cos 2(a-b)\big]$ in Eq.~(\ref{SQT1}) is replaced
by $\big[1+S_1S_2\cos 2(a-b)\big]$, see Fig.~\ref{fig3}(b).
\item
Replacing the rules Eq.~(\ref{SUBM1aa}) by the rules Eqs.~(\ref{SUBM1a}) and~(\ref{SUBM1b}),
and repeating the DES with the same value of $\alpha$ and $\beta$
also yields data (not shown) that are in excellent agreement with the quantum-theoretical description,
for $\gamma$ in the range $[0.1,0.98]$.
\item
Our DES model also reproduces the theoretical results (see Appendix~\ref{PROD}) if
the two photons of each pair have a fixed polarization (results not shown).
In Maxwell's theory, this case is described by light beams with fixed polarizations.
In quantum theory, this case is described by an (uncorrelated) product state.
\end{itemize}
Table~\ref{tab1} gives a compact overview of the agreemeent between the DES results
and the theoretical descriptions of the (E)EPRB experiments.

\begin{table*}[tbh]
\caption{Overview of the agreemeent between the DES results and the theoretical descriptions
of the (E)EPRB experiments.
MT: DES results agree with Maxwell's theory of electrodynamics;
QT: DES results agree with quantum theory of a pair of spin-1/2 particles.
Question mark: DES results cannot be described by
quantum theory of a pair of spin-1/2 particles (see Appendix~\ref{OT}).
}
\begin{ruledtabular}
\begin{tabular}{c@{\hspace{5mm}}|cc@{\hspace{5mm}}|cc} %
     & \multicolumn{2}{c@{\hspace{5mm}}|}{Without photon identification}&\multicolumn{2}{c}{With photon identification}\\
     \hline
     Photon pair polarization &  EPRB & EEPRB & EPRB & EEPRB \\
     \hline
     Orthogonal + random & MT [Eq.~(\ref{MAXW1}), $\phi_0=\pi/2$] & MT [Eq.~(\ref{MAXW2}), $\phi_0=\pi/2$] &
     QT [Eq.~(\ref{SQT0})] & QT [Eq.~(\ref{SQT1})] \\
     Parallel + random & MT [Eq.~(\ref{MAXW1}), $\phi_0=0$] & MT [Eq.~(\ref{MAXW2}), $\phi_0=0$] & ? & ? \\
     Fixed & MT [Eq.~(\ref{PROD0})] & MT [Eq.~(\ref{PROD1})] & QT [Eq.~(\ref{PROD0})] & QT [Eq.~(\ref{PROD1})] \\
   \end{tabular}
\end{ruledtabular}
\label{tab1}
\end{table*}

\section{Discussion and summary}

Laboratory EPRB experiments unavoidably require a procedure to classify a detection event
as corresponding to a photon or as something else.
Independent of the precise nature of this procedure (voltage threshold, local time window, time coincidence, etc.),
any model that aims at describing an EPRB experiment should, from the start, account for this procedure
by introducing additional variables into the description.
In contrast, Bell's model, while charmingly simple, does not account for an essential aspect
of laboratory EPRB experiments,
namely the classification of detection events in terms of photons or something else.
Consequently, any subquantum model that aims at reproducing the results of quantum theory
for the (E)EPRB experiment should have features that are not included in Bell's model.
As a matter of fact, a quick glance at how the data of laboratory EPRB experiments
are being processed reveals that it is the photon identification process which is lacking in Bell's model.
Including this process implies that correlations between events
are calculated only from subsets of the data, in which case Bell's theorem does not apply.
Using only subsets of the data, there is only the constraint that the correlation should, in absolute value,
be less than or equal to one. Apart from that ``almost everything'' is possible~\cite{PEAR70,PASC86,PASC87,BRAN87},
including a subquantum model that, in the appropriate limit, yields the correlation of the singlet state~\cite{RAED06c,ZHAO08}.

Clearly, on the basis of the statistical data alone, it is not possible to reject subquantum models of the EPRB and EEPRB experiments presented in this paper.
The relevant question is how a laboratory experiment can rule out or confirm
that (i) a subquantum level description is possible and/or (ii) the rules by
which the DES model of the beam splitter operates provide a reasonable description.

Regarding (i): If we consider it as irrelevant to ask what kind of process gives rise to the statistics of events,
it seems very difficult to beat quantum theory in terms of descriptive power~\cite{RAED14b}.
Therefore, it is clear that addressing (i) requires the analysis of the data on
the level of individual events, without being biased by what quantum theory predicts for the statistics.

Regarding (ii):
The DES model defined by Eqs.~(\ref{SUBM0}) and~(\ref{SUBM1aa}) or Eqs.~(\ref{SUBM0}), (\ref{SUBM1a}) and~(\ref{SUBM1b}) produces data in concert with Malus' law, i.e.~with the experiment,
and therefore seems solid.
The additional feature of the DES model (which allows us to reproduce the
statistics of the EPRB and EEPRB experiments as given by quantum theory)
is the subquantum law of retardation, defined by Eq.~(\ref{SUBM1aa}) or Eqs.~(\ref{SUBM1a}) and~(\ref{SUBM1b}).
At first sight, there is no experimental support for such a law.
However, let us look at the EPRB experiment from a slightly different perspective, namely,
as a setup to characterize the response of the observation stations to very feeble, randomly anticorrelated light.
Then, our DES data and also the analysis of experimental data~\cite{ZHAO08}
support the hypothesis that the EPRB experiment demonstrates that
the statistics of the detection events that have been classified as photons depend on the settings.

What if we try to measure the retardation by an experiment that uses feeble light with fixed polarization?
As explained in section~\ref{SUBM}, in our subquantum model the retardation time does not depend
on the setting of the beam splitter if the polarization is constant in time.
Only if the polarization is not fixed in time,
our subquantum model yields retardation times that depend on
the setting of the beam splitter.
This is precisely what the EPRB experiment does: through the correlation, it provides information about the retardation as a function of the setting of the beam splitter.
Therefore, to rule out the subquantum law of retardation used in our DES,
it is necessary to perform both the experiments with feeble, randomly polarized light and with feeble light of fixed polarization.

Summarizing, we have proposed a subquantum model which satisfies Einstein's criterion of locality and which generates, event-by-event, data that agrees with the quantum-theoretical
description of the Einstein-Podolsky-Rosen-Bohm and the extended  Einstein-Podolsky-Rosen-Bohm experiments.
This demonstration does not build on the traditional methods
of theoretical physics but instead uses a digital computer and a discrete-event simulation
as a metaphor for idealized, realizable laboratory experiments.

\appendix
\section{Quantum theory of the (E)EPRB experiment}\label{QT}

For reference, we briefly review the quantum-theoretical description of the EPRB and EEPRB experiment.
In this appendix, for the sake of generality, we first consider
magnetic spin-1/2 particles passing through Stern-Gerlach magnets.

In the context of (E)EPRB experiments, the case of interest is
a system of two spins in the singlet state, described by the density matrix
\begin{eqnarray}
\rho&=&
\frac{\openone-\bm\sigma_1\cdot\bm\sigma_2}{4}=
\frac{1}{2}\begin{pmatrix}
 \phantom{-}0 & \phantom{-}0 & \phantom{-}0 & \phantom{-}0 \\ 
 \phantom{-}0 & +1 & -1 & \phantom{-}0\\
 \phantom{-}0 & -1 & +1& \phantom{-}0\\
 \phantom{-}0 & \phantom{-}0\ & \phantom{-}0\ & \phantom{-}0\\
\end{pmatrix}=
\left( \frac{|{\uparrow}{\downarrow}\rangle-|{\downarrow}{\uparrow}\rangle}{\sqrt{2}}\right)
\left( \frac{\langle{\uparrow}{\downarrow}|-\langle{\downarrow}{\uparrow}|}{\sqrt{2}}\right)
.
\label{QT0}
\end{eqnarray}
A selective measurement on the spin-1/2 particle is described by the operator~\cite{BALL03}
\begin{eqnarray}
M(S,\bm\sigma,\bx)=\frac{\openone+S\;\bm\sigma\cdot\bx}{2}=M^2(S,\bm\sigma,\bx)
,
\label{P2D3}
\end{eqnarray}
projecting a state of the spin-1/2 system onto the eigenstate of $\bm\sigma\cdot\bx$ with eigenvalue $S=\pm1$.

The probabilities to observe the outcomes $S_1$ and $S_2$ in an EPRB experiment are given by~\cite{BALL03}
\begin{eqnarray}
P(S_1|\ba)&=&\mathbf{Tr}\; M(S_1,\bm\sigma_1,{\mathbf a})\;\rho\;M(S_1,\bm\sigma_1,{\mathbf a})=
\mathbf{Tr}\; \rho\;M(S_1,\bm\sigma_1,{\mathbf a})\;,
\nonumber \\
P(S_2|\bb)&=&\mathbf{Tr}\; M(S_2,\bm\sigma_2,{\mathbf b})\;\rho\;M(S_2,\bm\sigma_2,{\mathbf b})
=\mathbf{Tr}\; \;\rho\;M(S_2,\bm\sigma_2,{\mathbf b})\;
,
\label{QT2A}
\end{eqnarray}
respectively.
For two spin-1/2 particles in the singlet state Eq.~(\ref{QT0}), we have
$P(S_1|\ba)=\langle \bm\sigma_1\cdot\ba\rangle=1/2$ and $P(S_2|\bb)=\langle \bm\sigma_2\cdot\bb\rangle=1/2$,
The probability to observe the joint event $(S_1,S_2)$ is given by~\cite{BALL03}
\begin{eqnarray}
P(S_1,S_2|\ba,\bb)&=& 
\mathbf{Tr}\; M(S_2,\bm\sigma_2,{\mathbf b})M(S_1,\bm\sigma_1,{\mathbf a})\;
\rho\;M(S_1,\bm\sigma_1,{\mathbf a})M(S_2,\bm\sigma_2,{\mathbf b})
=\mathbf{Tr}\;
\rho\;M(S_1,\bm\sigma_1,{\mathbf a})M(S_2,\bm\sigma_2,{\mathbf b})
,
\label{QT2}
\end{eqnarray}
where we used the fact that $[M(S_1,\bm\sigma_1,\ba),M(S_2,\bm\sigma_2,\bb)]=0$ for all $\ba$ and $\bb$.
For two spin-1/2 particles in the singlet state Eq.~(\ref{QT0}), Eq.~(\ref{QT2}) becomes
\begin{eqnarray}
P(S_1,S_2|\ba,\bb)=\frac{1-S_1S_2\;\ba\cdot\bb}{4}
.
\label{QT2z}
\end{eqnarray}

Similarly, the probability to observe the joint event $(S_1,S_2,S_3,S_4)$ is given by
\begin{eqnarray}
P(S_1,S_2,S_3,S_4|\ba,\bb,\bc,\bd)
&=&\mathbf{Tr}\;\big[
M(S_4,\bm\sigma_2,{\mathbf d})M(S_3,\bm\sigma_1,{\mathbf c})M(S_2,\bm\sigma_2,{\mathbf b})M(S_1,\bm\sigma_1,{\mathbf a})
\nonumber \\
&&\hbox to 1cm{}
\rho\;
M(S_1,\bm\sigma_1,{\mathbf a})M(S_2,\bm\sigma_2,{\mathbf b})M(S_3,\bm\sigma_1,{\mathbf c})M(S_4,\bm\sigma_2,{\mathbf d})
\big]
\nonumber \\
&=&
\mathbf{Tr}\;\big[
\rho\;
M(S_1,\bm\sigma_1,{\mathbf a})
M(S_3,\bm\sigma_1,{\mathbf c})
M(S_1,\bm\sigma_1,{\mathbf a})
M(S_2,\bm\sigma_2,{\mathbf b})
M(S_4,\bm\sigma_2,{\mathbf d})
M(S_2,\bm\sigma_2,{\mathbf b})
\big]
.
\label{QT4}
\end{eqnarray}
Performing the matrix multiplications and calculating the trace we obtain
\begin{eqnarray}
P(S_1,S_2,S_3,S_4|\ba,\bb,\bc,\bd,Z)=\frac{1}{16}\big[1-S_1S_2\;\ba\cdot\bb\big]
\big[1+S_1S_3 \;\ba\cdot\bc\big]\big[1+S_2S_4\;\bb\cdot\bd\big]
\;.
\label{QT5}
\end{eqnarray}

The derivation of the quantum-theoretical description of an experiment with photon polarization instead of magnetic spin-1/2 particles is not much different, for details see Ref.~\onlinecite{RAED07a}.
The upshot is that we only have to replace $\ba\cdot\bb$ by $\cos2(a-b)$ etc.
Thus, in the case of an EEPRB experiment that
uses the polarization of the photons, the probability to observe the joint event $(S_1,S_2,S_3,S_4)$ is given by
\begin{eqnarray}
P(S_1,S_2,S_3,S_4|\ba,\bb,\bc,\bd,Z)=\frac{1}{16}\big[1-S_1S_2\cos 2(a-b)\big]
\big[1+S_1S_3 \cos2(a-c)\big]\big[1+S_2S_4\cos2(b-d)\big]
\;.
\label{QT6}
\end{eqnarray}

\section{A limitation of quantum theory for two spin-1/2 particles}\label{OT}
If the random polarizations of particles that enter BS1 and BS2 are the same instead
of orthogonal, the DES generates data which, within the usual statistical fluctuations,
is characterized by $E_1=E_2=0$, and $E_{12}=+\cos2(a-b)$.
In this appendix, we prove that a two-particle system with single-particle averages ${\widehat E}_1={\widehat E}_2=0$, and pair correlation ${\widehat E}_{12}=+\cos2(a-b)$ cannot be described by the
quantum theory of two spin-1/2 particles.
As in Appendix~\ref{QT}, we consider the general case of two spin-1/2 particles and deal with the case of photon polarization at the end.

Using the Pauli-matrices and the $2\times2$ unit matrix as a basis of the vector
space of $2\times2$ matrices, we can, without loss of generality,
write the $4\times4$ density matrix of the two-spin system as
\begin{eqnarray}
\widehat\rho=\frac{1}{4}\left(\openone
+ \sum_{k=x,y,z} u_k\sigma_1^k
+ \sum_{k=x,y,z} v_k\sigma_2^k
+ \sum_{k,l=x,y,z} \sigma_1^k w_{k,l}\sigma_2^l\right)
\;,
\label{OT0}
\end{eqnarray}
where the $u$'s and $v$'s are real numbers and the $w$'s are the elements of a Hermitian matrix.
According to quantum theory, we then have
\begin{eqnarray}
{\widehat E}_{1}&=&
\langle\bm\sigma_1\cdot\ba\rangle= \mathbf{Tr}\;\widehat\rho\;\bm\sigma_1\cdot\ba=\mathbf{u}\cdot\ba
\nonumber \\
{\widehat E}_{2}&=&\langle\bm\sigma_2\cdot\bb\rangle= \mathbf{Tr}\;\widehat\rho\;\bm\sigma_2\cdot\bb=\mathbf{v}\cdot\bb
\nonumber \\
{\widehat E}_{12}&=&
\langle\bm\sigma_1\cdot\ba\;\bm\sigma_2\cdot\bb\rangle= \mathbf{Tr}\;\widehat\rho\;\bm\sigma_1\cdot\ba\;\bm\sigma_2\cdot\bb
=\ba^{\mathrm{T}}\cdot\mathbf{w}\cdot\bb
\;.
\label{OT1}
\end{eqnarray}
If {\sl for all} unit vectors $\ba$ and $\bb$ we have
$\langle\bm\sigma_1\cdot\ba\rangle=\langle\bm\sigma_2\cdot\bb\rangle=0$
and $\langle\bm\sigma_1\cdot\ba\;\bm\sigma_2\cdot\bb\rangle=-q\,\ba\cdot\bb$,
then $\mathbf{u}=\mathbf{v}=0$ and $\mathbf{w}=-q\,\openone$, implying that
\begin{eqnarray}
\widehat\rho_{q}&=&
\frac{\openone-q\,\bm\sigma_1\cdot\bm\sigma_2}{4}
\;
.
\label{OT2}
\end{eqnarray}
The four eigenvalues of $\bm\sigma_1\cdot\bm\sigma_2$ are $1$, $1$, $1$, and $-3$.
Therefore, $\widehat\rho_{q}$ has
a negative eigenvalue if $q<-1/3$ and in this case $\widehat\rho_{q}$  does not qualify as
a density matrix whereas $\widehat\rho_{q=1}= (\openone-\bm\sigma_1\cdot\bm\sigma_2)/4$ does
(and represents the singlet state).

In summary, there does not exist a quantum-theoretical description in terms of a $4\times4$ density matrix
that yields ${\widehat E}_{1}={\widehat E}_{2}=0$
and ${\widehat E}_{12}=+|q|\ba\cdot\bb$
or for all unit vectors $\ba$ and $\bb$ and $|q|>1/3$.
This includes the special case for which ${\widehat E}_{12}=+\cos2(a-b)$ and ${\widehat E}_{1}={\widehat E}_{2}=0$.

\section{Product state}\label{PROD}

For completeness, we give the expressions for the probabilities for
the case that the two photons of each pair leave the source with fixed polarization $\bp$ and $\bq$,
respectively.
Instead of Eq.~(\ref{SQT0}), we have
\begin{eqnarray}
P(S_1,S_2|\ba,\bb,Z)=\frac{1+S_1\;\cos2(a-p)}{2} \frac{1+S_2\;\cos2(b-q)}{2}
,
\label{PROD0}
\end{eqnarray}
and instead of Eq.~(\ref{SQT1}), we have
\begin{eqnarray}
P(S_1,S_2,S_3,S_4|\ba,\bb,\bc,\bd,Z)=\frac{1}{16}\big[1+S_1\;\cos2(a-p)\big]\big[1+S_2\;\cos2(b-q)\big]
\big[1+S_1S_3 \cos2(a-c)\big]\big[1+S_2S_4\cos2(b-d)\big]
\;.
\nonumber \\
\label{PROD1}
\end{eqnarray}
Note that Eqs.~(\ref{PROD0}) and~(\ref{PROD1}) also apply to the case of classical optics with $I_0=1$.

\bibliography{../../../all20}
\end{document}